\documentclass[twoside,twocolumn,english,aps,showpacs,prl,superscriptaddress]{revtex4-1}
\usepackage[T1]{fontenc}
\usepackage{color}
\usepackage{bm}
\usepackage{amsmath}
\usepackage{amssymb}
\usepackage{graphicx}
\usepackage{esint}
\usepackage{mathrsfs}
\usepackage{booktabs}
\usepackage{multirow}
\usepackage{makecell}
\usepackage{verbatim}
\usepackage[plainpages = false, pdfpagelabels, 
                 bookmarks,
                 bookmarksopen = true,
                 bookmarksnumbered = true,
                 breaklinks = true,
                 linktocpage,
                 colorlinks = true,
                 linkcolor = blue,
                 urlcolor  = blue,
                 citecolor = blue,
                 anchorcolor = green,
                 hyperindex = true,
                 hyperfigures
                 ]{hyperref} 
\usepackage{dcolumn}   

\makeatletter
\newcommand*{\rom}[1]{\expandafter\@slowromancap\romannumeral #1@}
\makeatother

\usepackage{times}{\Large}
\newcommand{\ket}[1]{| #1 \rangle}
\newcommand{\bra}[1]{\langle #1 |}
\newcommand{\tr}{\textrm{Tr}}

\newcommand{\ketbra}[2]{| #1 \rangle\!\langle #2 |}

\newcommand{\dd}{\mathrm{d}}
\usepackage{tabularx}
\usepackage{amsthm}

\makeatother

\usepackage{babel}

\usepackage{lineno}

\begin{document}

\preprint{APS/123-QED}

\title{Mitigating Errors in Analog Quantum Simulation by Hamiltonian Reshaping or Hamiltonian Rescaling}

\author{Rui-Cheng Guo}
\affiliation{State Key Laboratory of Low Dimensional Quantum Physics, Department of Physics, Tsinghua University, Beijing, 100084, China}
\affiliation{Frontier Science Center for Quantum Information, Beijing 100184, China}

\author{Yanwu Gu}
\email{guyw@baqis.ac.cn}
\affiliation{Beijing Academy of Quantum Information Sciences, Beijing 100193, China}

\author{Dong E. Liu}
\email{dongeliu@mail.tsinghua.edu.cn}
\affiliation{State Key Laboratory of Low Dimensional Quantum Physics, Department of Physics, Tsinghua University, Beijing, 100084, China}
\affiliation{Frontier Science Center for Quantum Information, Beijing 100184, China}
\affiliation{Beijing Academy of Quantum Information Sciences, Beijing 100193, China}
\affiliation{Hefei National Laboratory, Hefei 230088, China}

\date{\today}

\begin{abstract}
  Simulating quantum many-body systems is crucial for advancing physics but poses substantial challenges for classical computers. Quantum simulations overcome these limitations, with analog simulators offering unique advantages over digital methods, such as lower systematic errors and reduced circuit depth, making them efficient for studying complex quantum phenomena. However, unlike their digital counterparts, analog quantum simulations face significant limitations due to the absence of effective error mitigation techniques. This work introduces two novel error mitigation strategies---Hamiltonian reshaping and Hamiltonian rescaling---in analog quantum simulation for tasks like eigen-energy evaluation. Hamiltonian reshaping uses random unitary transformations to generate new Hamiltonians with identical eigenvalues but varied eigenstates, allowing error reduction through averaging. Hamiltonian rescaling mitigates errors by comparing eigenvalue estimates from energy-scaled Hamiltonians. Numerical calculations validate both methods, demonstrating their significant practical effectiveness in enhancing the accuracy and reliability of analog quantum simulators.
\end{abstract}

\maketitle

\textbf{\large{{Introduction:}}}

Simulating quantum many-body Hamiltonians is a critical task for exploring complex phenomena in quantum physics. Key applications, including analyzing electronic structures in condensed matter physics~\cite{Bauer2020quantum}, as well as examining quantum phase transitions between many-body localization and thermalization~\cite{abanin2019mbl, Oganesyan2007mbl}, rely on accurately evaluating the eigen-energies of quantum many-body systems but pose significant challenges for classical computers.
Quantum computers hold the potential to efficiently solve these problems through various phase estimation algorithms \cite{nielsen&chuang,cleve1998quantum,kitaev1995quantum}. A more accessible version for the current noisy intermediate-scale quantum (NISQ) era \cite{preskillQuantumComputingNISQ2018} is control-free phase estimation \cite{kimmelRobustCalibrationUniversal2015,russoEvaluatingEnergyDifferences2021}, also known as the many-body spectroscopy technique \cite{jurcevic2015spectroscopy,roushanSpectroscopicSignaturesLocalization2017}. This technique determines the eigen-energies of a Hamiltonian by analyzing time series data of expectation values from time-evolved observables. The required Hamiltonian evolution can be implemented in both digital and analog quantum simulators \cite{daley2022practical,altman2021simulator}. Digital simulators, which are gate-based and capable of simulating universal Hamiltonians, face challenges such as systematic trotterization errors and increased complexity in circuit compilation. In contrast, analog simulators, tailored for specific tasks, are more straightforward and cleaner but offer limited programmability.

While digital and analog simulators offer distinct advantages, they both grapple with inherent challenges. A crucial concern for these platforms is the mitigation of the effect of noise and operational errors. Digital quantum simulators, combined with quantum error correction (QEC) \cite{shorSchemeReducingDecoherence1995,calderbankGoodQuantumErrorcorrecting1996,steaneErrorCorrectingCodes1996}, can achieve fault tolerance in the long term. However, implementing large-scale QEC remains challenging in the near term, despite some experiments reaching the break-even point \cite{google2023suppressing,ni2023beating,gupta2024encoding}. Analog quantum simulators, due to their limited controllability, lack effective methods to reduce errors. Therefore, developing suitable error cancellation methods for analog simulators, respecting their constraints, is of urgent importance.

One direct method to reduce errors is iterative calibration of the system Hamiltonian through Hamiltonian learning \cite{huangLearningManyBodyHamiltonians2023,gu2024practical} or quantum benchmarking methods \cite{choi2023preparing,mark2023benchmarking,gu2023benchmarking,shaw2024benchmarking} that are suitable for analog simulators. In addition to calibration, error can be mitigated at the circuit level through various quantum error mitigation (QEM) methods \cite{caiQuantumErrorMitigation2023}. QEM reduces errors by post-processing noisy data and requires significantly fewer quantum resources compared to quantum error correction. Despite the exponential sample scaling with system size \cite{takagiFundamentalLimitsQuantum2022,takagiUniversalSamplingLower2023,tsubouchiUniversalCostBound2023,quek2024exponentially}, QEM can enhance computational accuracy in the near term \cite{kandala2019error,kim2023evidence} and potentially reduce the resource overhead of QEC in the long term \cite{suzuki2022quantum}. However, most QEM methods are designed for digital quantum computers and require advanced spatial-temporal control. For example, probabilistic error cancellation \cite{temmeErrorMitigationShortDepth2017} necessitates first characterizing the noise channel of gates and then decomposing ideal gates into noisy ones. Virtual distillation requires many entangling gates between copies of system states \cite{huggins2021virtual,koczor2021exponential,liu2024virtual}. The noise resilient phase estimation in Ref.~\cite{guNoiseResilientPhaseEstimation2023} requires randomized compiling to tailor the noise into benign type for phase estimation. These methods are currently beyond the capabilities of analog quantum simulators, thus necessitating the development of tailored error mitigation techniques specifically for them.

In this work, we develop two error mitigation techniques for eigen-energy estimation in analog quantum simulators, termed Hamiltonian reshaping and Hamiltonian rescaling. We address the eigen-energy estimation problem using many-body spectroscopy techniques \cite{jurcevic2015spectroscopy,roushanSpectroscopicSignaturesLocalization2017}. Both procedures involve preparing an initial state, evolving it under the desired Hamiltonian at varied time intervals, and measuring the expectation values of pertinent observables. The eigen-energies are subsequently deduced from these expectation values using advanced signal processing techniques, such as the matrix pencil method \cite{sarkarUsingMatrixPencil1995}.

In the Hamiltonian reshaping approach, random unitary operations, ideally Pauli operations, are utilized to modify the original target Hamiltonian into alternative forms that retain the same eigen-energies but alter the eigenstates. By conducting many-body spectroscopy on each modified Hamiltonian, we derive several eigen-energy estimates. An average of these estimates provides an error-mitigated value up to the first order of noise strength.

In the Hamiltonian rescaling approach, we scale the original Hamiltonian by specific factors, resulting in new Hamiltonians that preserve the original eigenstates while altering the eigen-energies. By performing many-body spectroscopy with the rescaled Hamiltonians and adjusted time intervals, we effectively amplify the noise strength. Using the developed formulas, by combining the eigen-energy estimates from rescaled Hamiltonians, we achieve error mitigation up to the first or second order of noise intensity.

It is noteworthy  that our Hamiltonian rescaling method bears resemblance to error extrapolation techniques as detailed in \cite{temmeErrorMitigationShortDepth2017,liEfficientVariationalQuantum2017} during the data acquisition phase. However, there is a significant difference in the data processing approaches. Hamiltonian rescaling involves first obtaining eigen-energy estimates for each rescaled Hamiltonian, followed by error mitigation on these noisy estimates. In contrast, error extrapolation focuses on mitigating errors in the expectation values of observables at each time step prior to the extraction of eigen-energy estimates. Additionally, error extrapolation often relies on simplistic functional dependencies between expectation values and noise strength, such as polynomial or exponential relationships. These do not adequately capture the true relationship--a sum of complex exponential functions. As such, error extrapolation tends to offer less mitigation efficacy in this context compared to our Hamiltonian rescaling method, a conclusion supported by numerical experiments.
\\

\textbf{\large{{Results:}}}

\textbf{{\em Review of many-body spectroscopy technique.}}
Before discussing error mitigation, we first review a protocol designed to evaluate the eigen-energy differences of a given Hamiltonian using analog quantum simulators. This technique is commonly referred to as many-body spectroscopy \cite{jurcevic2015spectroscopy,roushanSpectroscopicSignaturesLocalization2017}. Additionally, this protocol has been extended to digital quantum simulators as robust phase estimation \cite{kimmelRobustCalibrationUniversal2015,russoEvaluatingEnergyDifferences2021}.

Consider an analog quantum simulator with \(n\) effective qubits, where the goal is to solve the eigenvalue problem of an effective Hamiltonian \(H\) that can be operated on the simulator. The eigenvalue problem of \(H\) is expressed as \(H \ket{\phi_j} = E_j \ket{\phi_j}\), with the set of eigenstates and corresponding eigenvalues denoted as \(\{\ket{\phi_j}, E_j\}\). Let \(S\) be the set of indices \(j\) such that \(j \in S\). Now, consider an arbitrary observable \(O\). By preparing the initial state \(\ket{\psi} = \sum_{j \in S} c_j \ket{\phi_j}\)
 and subsequently evolving it under the Hamiltonian 
\(H\) for a time duration 
\(t\), the expectation value \(\langle O \rangle(t)\) can be experimentally measured. The Fourier spectrum of \(\langle O \rangle(t)\)
 reveals information about the eigenvalue differences 
\(E_{l} - E_{m}\) (where \(l, m \in S\)). It can be demonstrated that the positions of the peaks in this spectrum correspond to these eigenvalue differences, thus enabling the solution of the eigenvalue problem. We proceed to present a detailed experimental protocol for general cases in four steps.

\begin{figure*}[t!]
  \centering
\includegraphics[width=\textwidth]{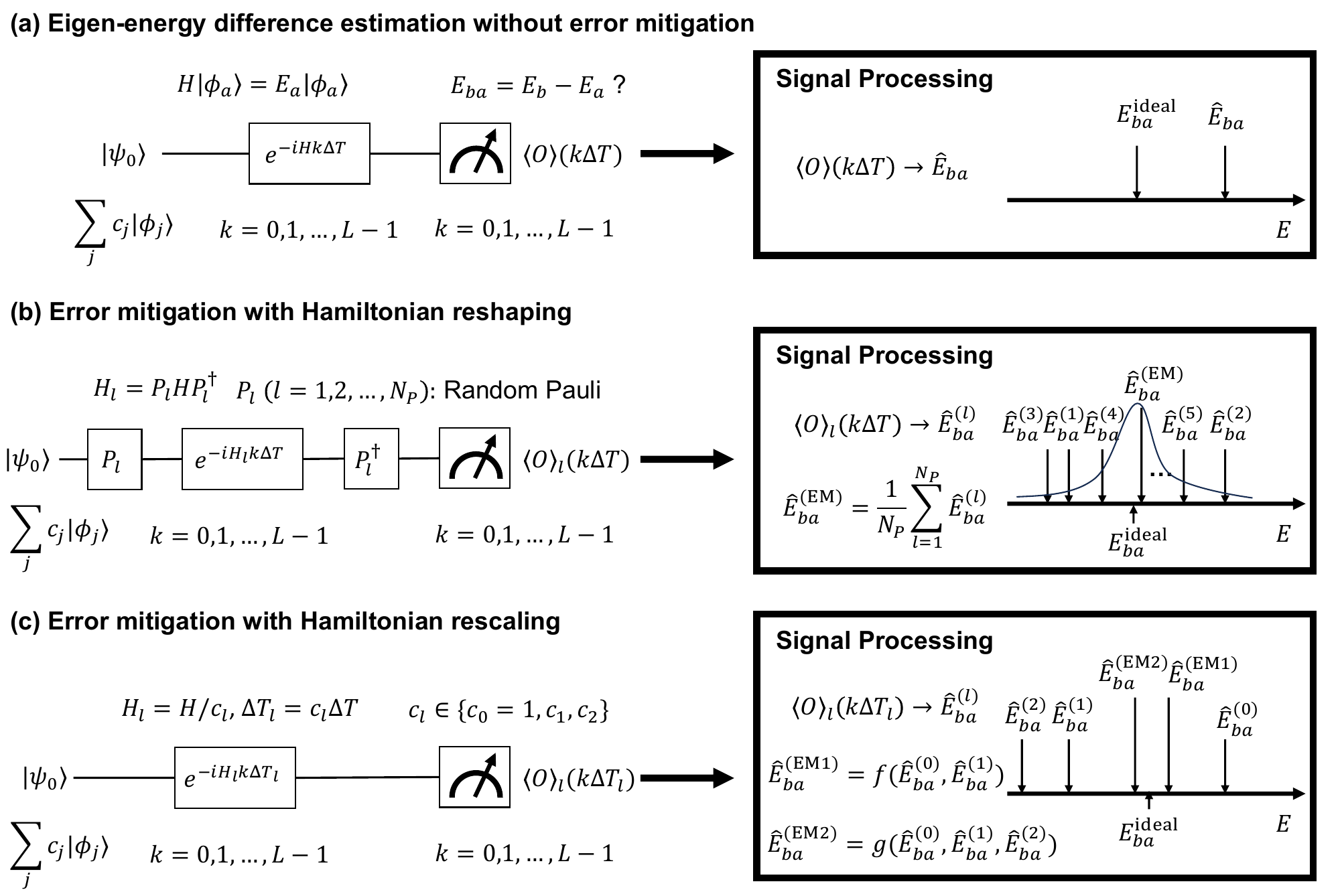}
  \caption{Noise mitigation techniques in many-body spectroscopy for eigen-energy estimation. (a) The many-body spectroscopy technique estimates the eigen-energy differences of the Hamiltonian $H$. A quantum state $\ket{\psi_0} = \sum_j c_j \ket{\phi_j}$ is prepared, representing a superposition of Hamiltonian eigenstates. The state is evolved over time intervals $k\Delta T$ (where $k = 0, 1, \ldots, L-1$), and the observable $O$ is measured to obtain the time-series signal $\{\langle O \rangle (k\Delta T)\}$. This signal is analyzed using methods like the matrix pencil method to determine the oscillating frequencies and corresponding eigen-energy differences. Noise in the Hamiltonian evolution causes deviations in the estimated energy differences $\hat{E}_{ba}$. We consider both the unitary noise caused by error Hamiltonian $H^{(1)}$ and the stochastic noise induced by the dissipator $\mathcal{D}[\cdot]$ as defined in Eq.~\eqref{eqn::errorTheory}. 
  (b) In Hamiltonian reshaping, random Pauli operations $P_l$ are applied to reshape the original Hamiltonian $H$ into new Hamiltonians $H_l = P_l H P_l^{\dagger}$. For each reshaped Hamiltonian $H_l$, the many-body spectroscopy technique is employed to estimate the eigen-energy difference $\hat{E}_{ba}^{(l)}$, with the initial state and measurement appropriately transformed. The averaged eigen-energy estimates $\hat{E}_{ba}^{(l)}$ from different reshaped Hamiltonians yield results that match the ideal values up to the first order of noise strength, assuming weak noise and first-order perturbation theory.
  (c) In Hamiltonian rescaling, the energy scale of the original Hamiltonian $H$ is adjusted to obtain new Hamiltonians $H_l = H/c_l$. For each rescaled Hamiltonian $H_l$, the many-body spectroscopy technique is used to estimate the eigen-energy difference $\hat{E}_{ba}^{(l)}$, with the rescaled time interval $\Delta T_l = c_l \Delta T$. By applying the developed formulas in Eq.~\eqref{eqn::First-Order} and Eq.~\eqref{eqn::secondOrderResult}, eigen-energy estimation errors can be mitigated up to the first or second order of noise strength using these $\hat{E}_{ba}^{(l)}$.
  }
\label{fig::protocol}

\end{figure*}

\begin{enumerate}
  \item Prepare an initial state \(\ket{\psi} = \sum_{j \in S} c_j \ket{\phi_j}\), which includes the eigenstates of interest.
  \item Evolve the quantum state with Hamiltonian \(H\) over different time: \(0\), \(\Delta T\), \(2\Delta T\), \(\dots\), \((L-1) \Delta T\), where we set time interval \(\Delta T < \frac{\pi}{\max (|E_{l} - E_{m}|)}\).
  \item Measure the observable \(O\) after each experiment to obtain time-series data \(\langle O \rangle (0)\), \(\langle O \rangle (\Delta T)\), \(\cdots\), \(\langle O \rangle ((L-1) \Delta T)\).
  \item Retrieve the information of eigen-energy differences  \(E_{ba} \Delta T\), where \(E_{ba} = E_b - E_a\) are the energy differences between \(\ket{\phi_b}\) and \(\ket{\phi_a}\), from time-series data \(\langle O \rangle (k\Delta T)\)  by signal processing methods, such as the matrix pencil method \cite{sarkarUsingMatrixPencil1995}.

\end{enumerate}
The protocol is illustrated in Fig.~\ref{fig::protocol}(a). The key steps of the proof are briefly summarized below.
The state at the time $k\Delta T$ is
    \begin{equation*}
        \ket{\psi(k\Delta T)}=\sum_{j\in S} c_j e^{-i k E_j \Delta T} \ket{\phi_j}.
    \end{equation*}
    Then the expectation value of the operator $O$ is
    \begin{align*}
        \langle O \rangle (k\Delta T)=\sum_{l \in S} \sum_{m \in S} c_{l} c_{m}^* \bra{\phi_{m}} O \ket{\phi_{l}}
        \cdot e^{-i k (E_{l}-E_{m}) \Delta T}.
    \end{align*}
    Thus, the expectation values $\langle O \rangle (k\Delta T)$ is a function of the variable $k$, which is composed of some oscillating functions. The oscillating frequencies $(E_{l}-E_{m}) \Delta T$ are relevant to the eigen-energy differences, which can be retrieved by some signal-processing methods.

The expectation value of the operator $O$ typically includes many eigen-energy differences, which complicates their retrieval. Therefore, it is necessary to limit the number of eigen-modes present in the data by selecting appropriate initial states and observables \cite{russoEvaluatingEnergyDifferences2021,guNoiseResilientPhaseEstimation2023,cuginiSpectralGapSuperposition2024}. Suppose we aim to evaluate the energy differences between the eigen-states \(\ket{\phi_b}\) and \(\ket{\phi_a}\). We can choose the initial state as \(\ket{\psi_0} = \frac{1}{\sqrt{2}} (\ket{\phi_a} + \ket{\phi_b})\) and a non-Hermitian measurement operator as \(2 \ketbra{\phi_b}{\phi_a}\), which can be decomposed into a complex superposition of Hermitian operators. In this configuration, the data \(\{\langle 2 \ketbra{\phi_b}{\phi_a} \rangle (k \Delta T)\}\) will only contain information about \(E_{ba} = E_b - E_a\), thereby facilitating the efficient retrieval of the specific energy difference. If one of the eigen-energies is known, the absolute value of the other eigen-energy can be computed. The state preparation and measurement process can be realized through adiabatic quantum evolution from an initial Hamiltonian, whose eigen-states are readily prepared, to the target Hamiltonian $H$, following the procedures outlined in \cite{cuginiSpectralGapSuperposition2024, robertsManybodyInterferometryQuantum2023, saxbergDisorderassistedAssemblyStrongly2022}. In addition to the adiabatic method, certain Hamiltonians offer natural choices for initial states and observables, such as excitation-conserving Hamiltonians. For these Hamiltonians, we can prepare the initial state as an equal superposition state of zero and single excitation states \cite{roushanSpectroscopicSignaturesLocalization2017,neillAccuratelyComputingElectronic2021,shiQuantumSimulationTopological2023,xiang2023simulating,wang2024fqh}. This approach ensures that the data will exclusively reflect the energy differences between the eigen-states in single excitation subspace and the zero excitation state.

\textbf{\em The effect of noise.}
In real quantum devices, noise is unavoidable and can affect the results of quantum algorithms. This section analyzes the impact of noise on many-body spectroscopy technique using matrix perturbation theory, under the assumption that the noise strength is weak relative to the energy scale of the target Hamiltonian. Additionally, it is assumed that the noise is time-independent.

Without loss of generality, we assume that the evolution of the noisy quantum simulator is described by the following Lindblad master equation:
\begin{equation}
    \frac{\dd}{\dd t}\rho(t) = -i[H_\text{exp}, \rho(t)] + \mathcal{D}[\rho(t)] \equiv \mathcal{L}[\rho(t)],
    \label{eqn::errorTheory}
\end{equation}
where $H_\text{exp} = H + H^{(1)}$ is the effective Hamiltonian in the experiment, $H^{(1)}$ represents the systematic error of the Hamiltonian (assumed to be small), $\mathcal{D}[\cdot] = \frac{1}{2} \sum_k (2L_k \cdot L_k^\dag - L_k^\dag L_k \cdot - \cdot L_k^\dag L_k)$ is the dissipator ($L_k$ are Lindblad operators), and $\mathcal{L}$ is the Liouvillian super-operator. Denote $\widetilde{\mathcal{D}}[\cdot] = -i[H^{(1)}, \cdot] + \mathcal{D}[\cdot]$ as the noise super-operator such that
\begin{equation}
    \frac{\dd}{\dd t}\rho(t) = -i[H, \rho(t)] + \widetilde{\mathcal{D}}[\rho(t)].
\end{equation}
Consider the eigenvalue problem of the Liouvillian super-operator. If the noise strength is weak ($|\widetilde{\mathcal{D}}| \ll |[H, \cdot]|$), $\widetilde{\mathcal{D}}$ can be considered as a small perturbation. The right eigenvalues and eigen-states of the Liouvillian super-operator before perturbation are the same as those of the super-operator $-i[H, \cdot]$, which are $\{i(E_b - E_a)\}$ and $\{\ketbra{\phi_a}{\phi_b}\}$.
Consider the non-degenerate first-order perturbation theory, the eigenvalues of the Liouvillian super-operator are
\begin{equation}
    \Lambda_{ab} \approx i(E_b-E_a)+\bra{\phi_a}\widetilde{\mathcal{D}}[\ketbra{\phi_a}{\phi_b}]\ket{\phi_b}
\end{equation}
with the corresponding eigen-operators $M_{ab}$ such that $\mathcal{L}[M_{ab}]=\Lambda_{ab}M_{ab}$.
The density matrix can be expanded by the set of eigen-operators $M_{ab}$
\begin{equation}
    \rho(t)=\sum_{ab} c_{ab}(t) M_{ab}\,.
    \label{eqn::cab}
\end{equation}
Substituting the equation above into the master equation we obtain
\begin{equation}
    \frac{\dd}{\dd t}\sum_{ab} c_{ab}(t) M_{ab}=\sum_{ab} \Lambda_{ab} c_{ab}(t) M_{ab},
\end{equation}
so that
\begin{equation}
    c_{ab}(t)
    = c_{ab}(0) e^{\Lambda_{ab} t},
    \label{eqn:cabt}
\end{equation}
where the initial time is set as zero.
Then the expectation value of the operator $O$ after time $k \Delta T$ is
\begin{eqnarray}\label{eqn::DampedOscillation}
    \langle O \rangle (k\Delta T) &=& \tr (O \rho(k\Delta T))\nonumber\\
    &=&\sum_{ab} c_{ab}(0) \tr(O M_{ab}) e^{\Lambda_{ab}k\Delta T}\nonumber\\
    &=& \sum_{ab} C_{ab} r_{ab}^k e^{i k \delta \phi_{ab}}
\end{eqnarray}
where we define
\begin{eqnarray}\label{eqn::Cab}
    r_{ab}&=&e^{\mathrm{Re} \bra{\phi_a}\widetilde{\mathcal{D}}[\ketbra{\phi_a}{\phi_b}]\ket{\phi_b} \Delta T},\\
    C_{ab}&=&c_{ab}(0) \tr(O M_{ab})
\end{eqnarray}
and
\begin{equation}
    \delta \phi_{ab}=(E_b-E_a) \Delta T + \mathrm{Im} \bra{\phi_a}\widetilde{\mathcal{D}} [\ketbra{\phi_a}{\phi_b}] \ket{\phi_b} \Delta T.   \label{eqn::deltaPhiAB}
\end{equation}
This result indicates that expectation value $\langle O \rangle (k\Delta T)$ can be expressed as a sum of damped oscillation modes. Signal processing techniques can still be employed to extract the phase information $\delta\phi_{ab}$, thereby allowing for the estimation of the eigen-energy difference $E_{ba}=E_b-E_a$. This protocol is resilient to state preparation and measurement (SPAM) errors, as these errors affect only the value of $C_{ab}$ and do not influence the phase $\delta \phi_{ab}$.

However, from Eq.~\eqref{eqn::deltaPhiAB}, we observe that the eigen-energy estimate $\hat{E}_{ba} = \delta \hat{\phi}_{ab}/\Delta T$ is biased due to noise, unless $\bra{\phi_a}\widetilde{\mathcal{D}}[\ketbra{\phi_a}{\phi_b}]\ket{\phi_b} \in \mathbb{R}$. A sufficient condition for an unbiased estimate, up to the first order of noise strength, is that the Lindblad operators $L_k$ are Hermitian and that the systematic error $H^{(1)} = 0$. This condition reflects the requirements outlined in our previous work \cite{guNoiseResilientPhaseEstimation2023}. However, Ref.~\cite{guNoiseResilientPhaseEstimation2023} necessitates randomized compiling to tailor general noise to the desired type, a technique that is challenging to implement in analog quantum simulators due to their limited control capabilities. Therefore, there is a need to develop appropriate error mitigation methods specifically tailored for analog simulators.

We now introduce two error mitigation strategies tailored for the Hamiltonian eigen-energy estimation problem in analog quantum simulators: \textit{Hamiltonian reshaping} and \textit{Hamiltonian rescaling}.

\textbf{{\em Hamiltonian reshaping strategy.}}
The Hamiltonian reshaping technique involves applying random unitary transformations to the original target Hamiltonian, generating new Hamiltonians that retain the same eigen-energies while possessing different eigenstates. This approach allows the eigen-energy estimates derived from these transformed Hamiltonians to be influenced by distinct aspects of noise. We show that averaging these estimates leads to an unbiased eigen-energy estimate, accurate up to the first order in noise strength.

Consider a random unitary operation $U$ from a set $G$, which transforms the Hamiltonian $H$ into $H_U = U H U^\dagger$. The eigenvalues of $H_U$ remain identical to those of $H$, while its eigenstates transform to $U|\phi_j\rangle$. Using the many-body spectroscopy technique in the previous section, we can estimate the phase $\delta \phi_{ab}^{(U)}$ in the time interval $\Delta T$ for $H_U$ between the states $U|\phi_b\rangle$ and $U|\phi_a\rangle$
\begin{equation}
    \delta \phi_{ab}^{(U)}=E_{ba}\Delta T + \mathrm{Im} \bra{\phi_a}U^\dagger \widetilde{\mathcal{D}} [U\ketbra{\phi_a}{\phi_b}U^\dagger] U \ket{\phi_b} \Delta T.
\end{equation}
The eigen-energy estimate $\hat{E}_{ba}^{(U)}=\delta \hat{\phi}_{ab}^{(U)}/\Delta T$  for each Hamiltonian $H_U$ is biased.
However, if we average these eigen-energy estimates from different $H_U$ with $U\in G$, then
\begin{equation}
    \overline{\delta \phi_{ab}^{(U)}}=E_{ba}\Delta T + \overline{\mathrm{Im} \bra{\phi_a}U^\dagger \widetilde{\mathcal{D}} [U\ketbra{\phi_a}{\phi_b}U^\dagger] U \ket{\phi_b}} \Delta T.
    \label{eqn::Strategy1}
\end{equation}
We can design a set $G$ to ensure the condition
\begin{equation}
    \overline{\mathrm{Im} \bra{\phi_a}U^\dagger \widetilde{\mathcal{D}} [U\ketbra{\phi_a}{\phi_b}U^\dagger] U \ket{\phi_b}}=0
    \label{eqn::condition}
\end{equation}
and in this case, the estimated result of $E_{ba}$ after mitigation is
\begin{equation}
    \hat{E}_{ba}^{(\text{EM})}=\overline{\delta \hat{\phi}_{ab}^{(U)}}/\Delta T=\overline{\hat{E}_{ba}^{(U)}}=E_{ba}
    \label{eqn::ReshapingTheory}
\end{equation}
which is unbiased up to the first order of noise strength.

There are numerous sets $G$ that satisfy the condition outlined in Eq.~(\ref{eqn::condition}), including the Clifford group and the Pauli group. However, when the Hamiltonian is expanded in the Pauli basis, the Clifford-transformed Hamiltonians exhibit Pauli terms that differ significantly from those of the original Hamiltonian, thereby complicating practical experimental implementation. Conversely, a Pauli-transformed Hamiltonian retains the same Pauli terms as the original Hamiltonian, albeit with opposite signs sometimes. Utilizing the properties of Pauli twirling \cite{caiConstructingSmallerPauli2019}, we demonstrate that if $G$ comprises the set of $n$-qubit Pauli operations and $H$ is a Hamiltonian in an $n$-qubit quantum system, the condition necessary for error mitigation, as specified in Eq.~(\ref{eqn::condition}), is satisfied for a general noise channel $\widetilde{\mathcal{D}}$ (see ``Methods''). We refer to this method as Hamiltonian reshaping, which is also employed in Hamiltonian learning \cite{huangLearningManyBodyHamiltonians2023} and noise tailoring \cite{santos2024pseudo}. This approach is illustrated in Fig.~\ref{fig::protocol}(b).

\begin{figure*}[t!]
  \centering
  \includegraphics[width=\textwidth]{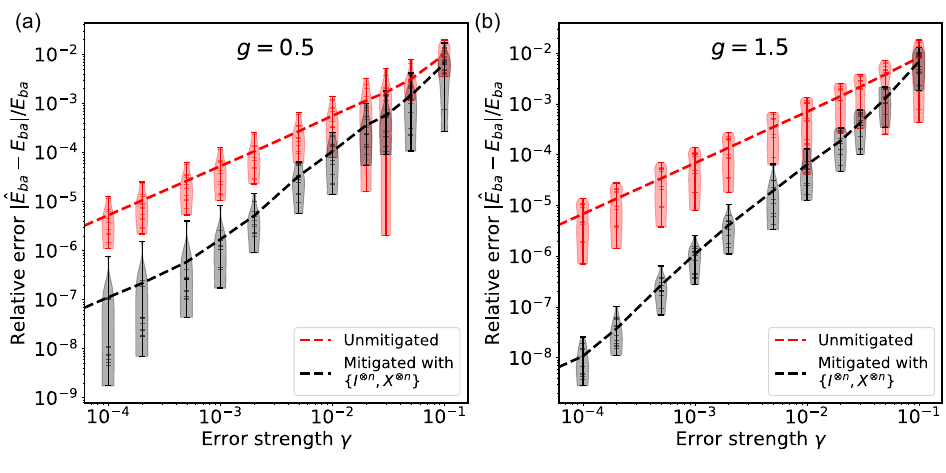}
  \caption{Effect of T1 noise and the efficacy of Hamiltonian reshaping tailored for T1 noise. The figure demonstrates the impact of T1-type noise on a model Hamiltonian $H = -\sum_i X_i - \sum_i Y_i - g \sum_{\langle i,j \rangle} X_i X_j$, with systematic error $H^{(1)} = \kappa \beta \sum_{j=1}^n Z_j$, and the effectiveness of a Hamiltonian reshaping technique. T1 noise is modeled by the Lindblad operator $L_k = \sqrt{\kappa} \ket{0}\bra{1}_k \otimes \mathbf{I}_{\text{others}}$, where $k=1, \ldots, n$ and $n=6$ qubits. The relative error strength is $\gamma = \kappa/|E_{ba}|$. Parameters are set as $\beta = 0.01$, $\Delta T = 0.0001$, $L = 2000$.
  Panels (a) and (b) illustrate that the estimation of the energy difference $E_{ba}$
is affected by noise in a manner that varies with the interaction strength $g$. The Hamiltonian reshaping method using $\{\mathbf{I}^{\otimes n}, \mathbf{X}^{\otimes n}\}$ 
provides effective mitigation under these conditions.}
  \label{fig::T1noise}
\end{figure*}

The Hamiltonian reshaping can be effectively implemented in typical analog quantum simulators by selecting an appropriate set of random unitary operators that align with the constraints of the target system. As previously mentioned, the preferred choice for $G$ is a set of Pauli operations, which only alters the sign of Pauli terms in a Hamiltonian. Generally, the sign of local Pauli operators can be altered by adjusting the detuning and phase of the control pulse. The interaction Hamiltonian terms often consist of weight-2 Pauli operators, such as 
$X\otimes X$, $Z\otimes Z$ and exchange interaction $X\otimes X+Y\otimes Y$. Various techniques are available for modifying their signs in superconducting circuits \cite{chen2014tunable,yanTunableCouplingScheme2018,cai2019observation}, trapped ions \cite{sorensenQuantumComputationIons1999,zhangObservationManybodyDynamical2017,monroeProgrammableQuantumSimulations2021}, and neutral atoms \cite{Saffman_2016,browaeys2020many,morgado2021quantum,sylvain2019observation}. However, it is important to note that certain Pauli operators in the protocol transform the exchange interaction $X\otimes X+Y\otimes Y$ to $X\otimes X-Y\otimes Y$, which is usually unattainable in experimental settings. In such cases, it is necessary to select a smaller set of Pauli operators rather than the full Pauli group to reshape the Hamiltonian while preserving the interaction terms. A particular choice is the set $\{I^{\otimes n}, X^{\otimes n}, Y^{\otimes n}, Z^{\otimes n}\}$, where each element commutes with the exchange interactions, thereby preserving these interactions. Although this smaller set of Pauli operators $\{I^{\otimes n}, X^{\otimes n}, Y^{\otimes n}, Z^{\otimes n}\}$ may not fully mitigate errors arising from general noise, it can mitigate errors due to any local noise.  Moreover, reshaping with the full Pauli group suffers from statistical errors due to the limited samples of Pauli operations. Consequently, the smaller set $\{I^{\otimes n}, X^{\otimes n}, Y^{\otimes n}, Z^{\otimes n}\}$ performs better when dealing with local noise. An even smaller set, 
$\{I^{\otimes n}, X^{\otimes n}\}$, is effective in mitigating errors due to amplitude damping noise and local $Z$ unitary errors, which are the most prevalent types of noise in quantum systems, as illustrated in Fig.~\ref{fig::T1noise}.

We summarize the Hamiltonian reshaping strategy as follows:
\begin{itemize}
    \item Step 1: Generate a set of random Hamiltonians $H_U = U H U^\dagger$, reshaped by a random unitary operation $U \in G$. Here, $G$ is preferably chosen as a set of Pauli operations.
    \item Step 2: For each reshaped Hamiltonian $H_U$, estimate $\hat{E}_{ba}^{(U)}$ using the many-body spectroscopy technique.
    \item Step 3:
    Compute the error mitigated estimate of energy difference $E_{ba}$ by applying Eq.~\eqref{eqn::ReshapingTheory}.
\end{itemize}

\textbf{{\em Hamiltonian rescaling strategy.}}
The Hamiltonian rescaling method, on the other hand, involves adjusting the energy scale of the target Hamiltonian, producing a series of Hamiltonians with identical eigenstates but differing eigen-energies. Correspondingly, we modify the time scale to preserve the ideal phase evolution at each time interval, effectively amplifying the impact of noise on the eigen-energy estimates. By combining the estimates from both the original and rescaled Hamiltonians using appropriately developed formulas, we can mitigate errors up to the first or second order in noise strength.

Inspired by Ref.~\cite{temmeErrorMitigationShortDepth2017}, we develop an error mitigation strategy based on the concept of Hamiltonian rescaling. 
Notably, our approach distinctly diverges from the conventional zero-noise extrapolation methods described in \cite{liEfficientVariationalQuantum2017,temmeErrorMitigationShortDepth2017}, aiming to offer a unique perspective on reducing quantum errors.

First, we obtain the estimate $\hat{E}_{ba}^{(0)}=\hat{E}_{ba}=\delta \hat{\phi}_{ab}/\Delta T$ from the original Hamiltonian using the many-body spectroscopy technique. Next, we consider the rescaled Hamiltonian $H'(t)= \frac{1}{c} H(t/c)$ where $c \in (1, +\infty)$, let $\Delta T' = c \Delta T$, and obtain the corresponding $\delta \phi_{ab}'$ using the same protocol. Thus, we have
\begin{align}
    \delta \phi_{ab}' = \frac{1}{c} E_{ba} \Delta T' + \mathrm{Im} \langle \phi_a | \widetilde{\mathcal{D}} [|\phi_a \rangle \langle \phi_b |] |\phi_b \rangle \Delta T'.
    \label{eqn::deltaPhiABprime}
\end{align}
Here, we assume the dissipator is time-translation invariant, so that the noise effect does not change if we only rescale the strength of the Hamiltonian. This assumption, also made in \cite{temmeErrorMitigationShortDepth2017, kandalaExtendingComputationalReach2019}, is crucial for experimentally controlling the noise strength in the zero-noise extrapolation method \cite{temmeErrorMitigationShortDepth2017, liEfficientVariationalQuantum2017}.
Nevertheless, our objective differs from this approach, as we aim to directly counteract the effects of noise rather than merely extrapolating to a zero-noise scenario. Although the fluctuations of $T_1$ and $T_2$ errors exist over large time scales in real experiments, if measurements for different rescaling factors for the same evolution time are performed in quick succession, this assumption can be justified according to the experiment in \cite{kandalaExtendingComputationalReach2019}.

We define $\hat{E}_{ba}^{(1)} = \delta \hat{\phi}_{ab}' / \Delta T'$ as a biased estimator for $E_{ba}/c$. Combining equations \eqref{eqn::deltaPhiAB} and \eqref{eqn::deltaPhiABprime}, we obtain the error mitigated energy estimate
\begin{equation}
\hat{E}_{ba}^{(\text{EM1})} = \frac{c}{c-1} (\frac{\delta \phi_{ab}}{\Delta T} - \frac{\delta \phi_{ab}^\prime}{\Delta T^\prime})=\frac{c}{c-1} (\hat{E}_{ba}^{(0)}-\hat{E}_{ba}^{(1)}).
    \label{eqn::First-Order}
\end{equation}
It should be noted that the above equation does not explicitly depend on noise factors; hence, this strategy effectively mitigates the impact of noise. It is important to note that the parameter $c$ must not be excessively large, as $\widetilde{\mathcal{D}}[\cdot]$ is intended to act as a minor perturbation in comparison to $-i\left[\frac{1}{c}H(t/c),\cdot\right]$, ensuring the effective application of perturbation theory. Furthermore, a larger $c$ reduces the estimation error of $E_{ba}$, as this is directly proportional to $\frac{c}{c-1}$. 

We now examine the second-order corrections to the non-degenerate eigenvalues of the Liouvillian operator $\mathcal{L}$. The Liouvillian is defined as $\mathcal{L}[\cdot]=-i[H,\cdot]+\widetilde{\mathcal{D}}[\cdot]$, with the left and right eigenvectors of the unperturbed part $-i[H,\cdot]$ being $\{\ketbra{\phi_a}{\phi_b}\}$, corresponding to eigenvalues $\{iE_{ba}\}$. The outcomes of the second-order corrections can be derived using eigenvalue perturbation theory~\cite{bamiehTutorialMatrixPerturbation2022}:
\begin{align}
    &\Lambda_{ab} \approx \Lambda_{ab}^{(0)}+\Lambda_{ab}^{(1)}+\Lambda_{ab}^{(2)}\nonumber\\
    &= iE_{ba} + \bra{\phi_a}\widetilde{\mathcal{D}}[\ketbra{\phi_a}{\phi_b}]\ket{\phi_b}\nonumber\\
    &+\sum_{(p,m)\neq (a,b)} \frac{\bra{\phi_p}\widetilde{\mathcal{D}}^{\dagger}[\ketbra{\phi_a}{\phi_b}]\ket{\phi_m}\bra{\phi_a}\widetilde{\mathcal{D}}[\ketbra{\phi_p}{\phi_m}]\ket{\phi_b}}{i(E_{ba}-E_{mp})}.
    \label{eqn::secondOrder}
\end{align}
 To effectively implement error mitigation, at least two rescaling factors, $c_1$ and $c_2$, are required. Under this scheme, we can extract estimates $\hat{E}_{ba}^{(l)}=\delta \hat{\phi}_{ab}^{(l)}/ \Delta T_l$ by utilizing the rescaled Hamiltonian $H_l=H/c_l$, where the corresponding time interval is given by $\Delta T_l=c_l \Delta T$. As
\begin{align}
    &\delta \phi_{ab}^{(l)}/ \Delta T_l =\frac{1}{c_l} E_{ba} + \mathrm{Im} \bra{\phi_a}\widetilde{\mathcal{D}}[\ketbra{\phi_a}{\phi_b}]\ket{\phi_b} \nonumber\\
    &+c_l \sum_{(p,m)\neq (a,b)} \mathrm{Im} \frac{\bra{\phi_p}\widetilde{\mathcal{D}}^{\dagger}[\ketbra{\phi_a}{\phi_b}]\ket{\phi_m}\bra{\phi_a}\widetilde{\mathcal{D}}[\ketbra{\phi_p}{\phi_m}]\ket{\phi_b}}{i(E_{ba}-E_{mp})},
\end{align}
we can mitigate the error up to the second order of noise strength by
\begin{align}
&\hat{E}_{ba}^{(\text{EM2})} = \nonumber\\
    &\frac{c_1 c_2 \left((1-c_2)(\hat{E}_{ba}^{(1)} - \hat{E}_{ba}^{(0)})+(c_1-1)(\hat{E}_{ba}^{(2)} - \hat{E}_{ba}^{(0)})\right)}{(c_2-c_1)(c_1-1)(c_2-1)}.
    \label{eqn::secondOrderResult}
\end{align}

\begin{figure}[t!]
    \centering
    \includegraphics[width=1\columnwidth]{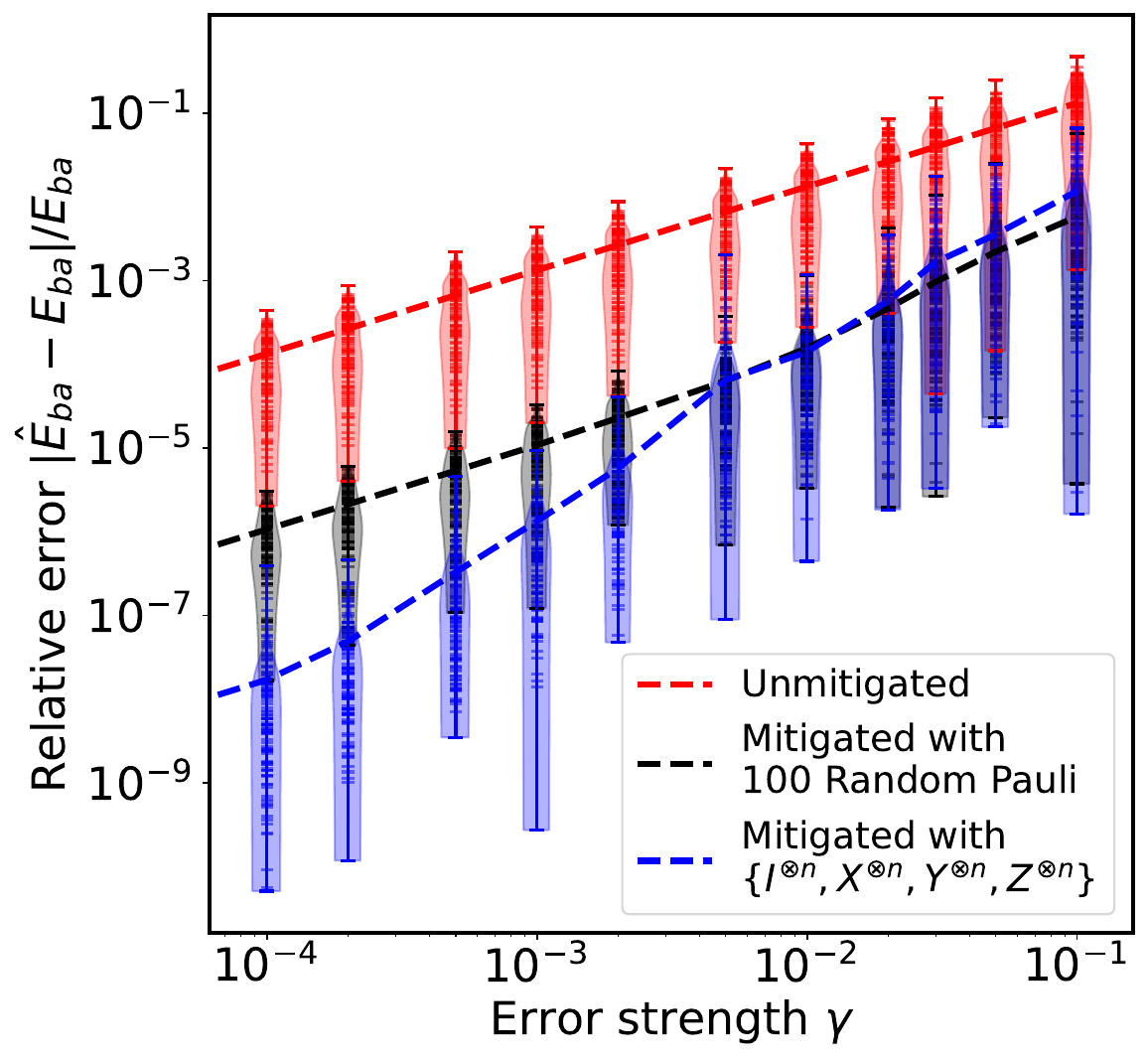}
    \caption{Numerical results of Hamiltonian reshaping. The eigen-energy differences $E_{ba} = E_b - E_a$ for 100 randomly chosen pairs of eigenstates $(a,b)$ are estimated. We consider the Hamiltonian in Eq.~\eqref{eqn::simulationHamiltonian} with parameters $n=6$, $\nu_z=4$, $\nu_x=1$, and $J=4$. The Lindblad operators are assumed to be $L_k = \sqrt{\kappa} (i\ket{0}\bra{0}_k + \ket{1}\bra{1}_k) \otimes \mathbf{I}_{\text{others}}$ for $k = 1, \ldots, n$, and the error Hamiltonian is $H^{(1)} = \kappa \beta \sum_{j=1}^n Z_j$. Parameters are set to $\beta = 0.01$, $\Delta T = 0.0001$, and $L = 2000$. For each numerical experiment, $\kappa = \gamma |E_{ba}|$ is used to reflect the relative error strength. The red line represents the average estimated relative error before error mitigation. Error mitigation is performed using 100 Pauli reshaped Hamiltonians, with the black line showing the average result after mitigation. The blue line indicates the mitigation result using reshaped Hamiltonians with $\{I^{\otimes n},X^{\otimes n},Y^{\otimes n},Z^{\otimes n}\}$. When the error rate is small, this method performs better due to reduced statistical errors, as discussed in ``Results''. The violin plot illustrates the distribution of relative errors before or after mitigation for these 100 $E_{ba}$ values.
    }
    \label{fig::randomPauli}
\end{figure}

We note that this Hamiltonian rescaling strategy is different from the Richardson extrapolation method in \cite{temmeErrorMitigationShortDepth2017}.
They both use the experiment data before and after rescaling the Hamiltonian, but process the experiment data with different methods (see Fig.~\ref{fig::CompareWithRE}(a)).
Our Hamiltonian rescaling strategy is designed to address the damped oscillation relationship described by Eq.~\eqref{eqn::DampedOscillation}, a feature that the Richardson extrapolation method cannot fully capture. Consequently, our approach provides a more customized solution for addressing the complexities intrinsic to the many-body spectroscopy problem. Furthermore, this strategy is experimentally viable, as the coefficients of the effective Hamiltonian can be modified using the same method employed in Hamiltonian reshaping.

We summarize this strategy into three steps:
\begin{itemize}
    \item Step 1: Choose \(c_1, c_2 > 1\).
    \item Step 2: Estimate \(\hat{E}_{ba}^{(0)} = \frac{\delta \hat{\phi}_{ab}}{\Delta T}\), \(\hat{E}_{ba}^{(1)} = \frac{\delta \hat{\phi}_{ab}^{(1)}}{\Delta T_1}\), and \(\hat{E}_{ba}^{(2)} = \frac{\delta \hat{\phi}_{ab}^{(2)}}{\Delta T_2}\) by many-body spectroscopy technique with the original and rescaled Hamiltonian \(H(t) = H\), \(H_1 (t) = \frac{1}{c_1} H\left(\frac{t}{c_1}\right)\), and \(H_2 (t) = \frac{1}{c_2} H\left(\frac{t}{c_2}\right)\).
    \item Step 3: Mitigate the energy estimation error by Eq.~\eqref{eqn::secondOrderResult}.
\end{itemize}

\begin{figure}[t!]
  \centering
  \includegraphics[width=1\columnwidth]{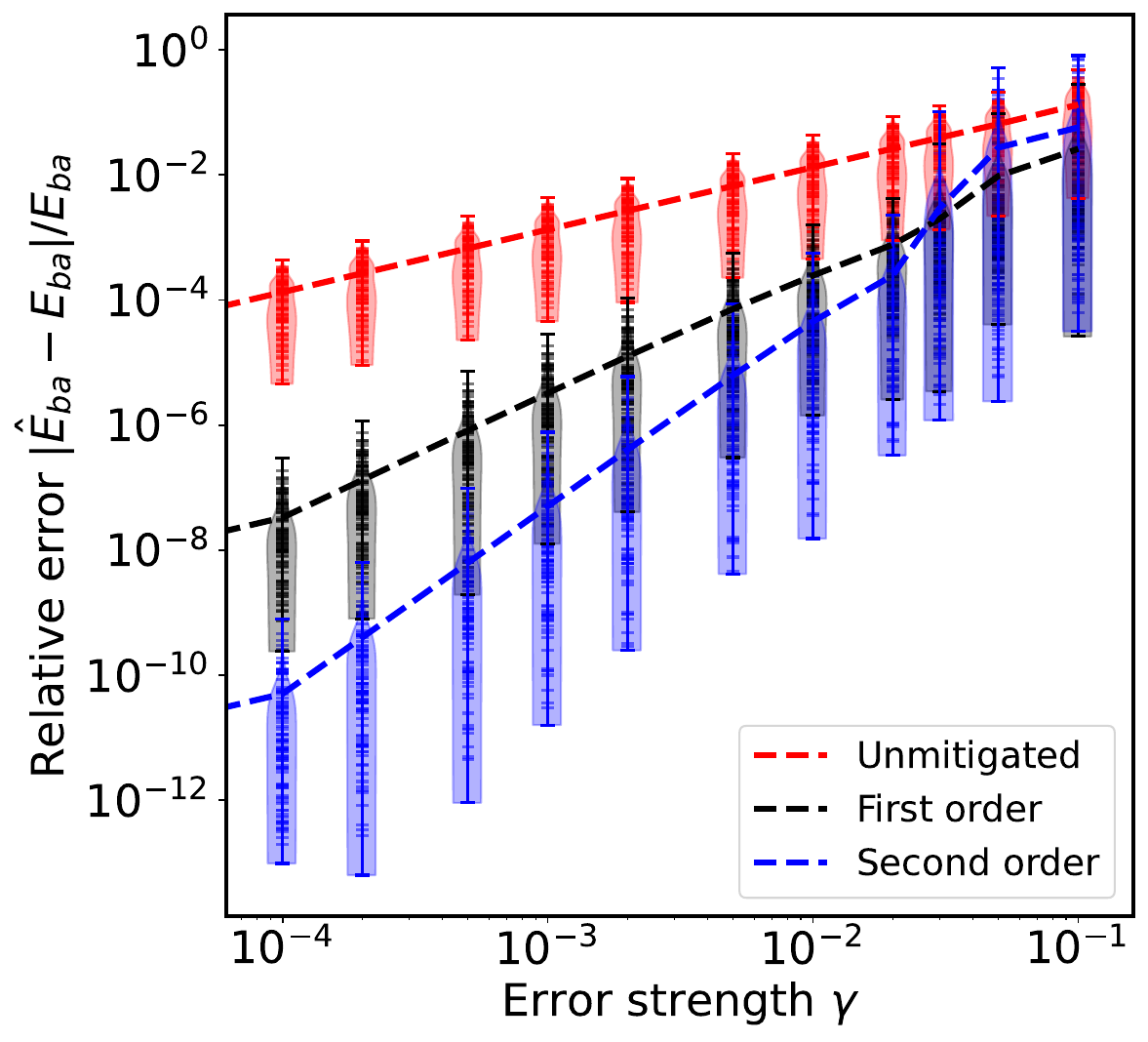}
  \caption{Numerical results of Hamiltonian rescaling. The same 100 pairs of eigen-states $(a,b)$ as those in Fig.~\ref{fig::randomPauli} are considered.
  We choose scale factors $c_1=2$ and $c_2=1.5$.
  The other settings are the same as that in Fig.~\ref{fig::randomPauli}.
  The dashed lines show the relative error of these 100 eigenvalue differences in average, and the violin plot shows their corresponding distribution. These results demonstrate that our strategy significantly mitigates the relative error, with the second-order correction providing superior accuracy compared to the first-order correction. In the log-log scale plot, the slope of the unmitigated line is approximately 1, consistent with Eq.~\eqref{eqn::deltaPhiAB}. The slope of the first-order correction line is close to 2, and that of the second-order correction is approximately 3, indicating effective mitigation of errors up to the first and second orders.}
  \label{fig::rescaling}
\end{figure}

\textbf{{\em Numerical results.}}
We demonstrate our error mitigation strategies with numerical experiments, performed using the simulation package QuTiP \cite{johanssonQuTiPOpensourcePython2012, johanssonQuTiPPythonFramework2013}. We consider a Hamiltonian defined on a qubit ring model, which is a modification from the model discussed in \cite{SolvingLindbladDynamics}:
\begin{equation}
    H = \frac{1}{2} \sum_{i=0}^{n-1} (2\pi \nu_z Z_i + 2\pi \nu_x X_i) + \frac{1}{2} \sum_{\langle i,j \rangle} 2\pi J (X_i X_j + Y_i Y_j).
    \label{eqn::simulationHamiltonian}
\end{equation}
Here, \(\langle i,j \rangle\) indicates that \(i\) and \(j\) are the nearest neighbors on a qubit ring. The dissipator of the dynamics is given by
\begin{equation}
    \mathcal{D}[\rho(t)] = \sum_k \frac{1}{2} \left(2 L_k \rho(t) L_k^\dagger - \rho(t) L_k^\dagger L_k - L_k^\dagger L_k \rho(t)\right)
\end{equation}
where $L_k = \sqrt{\kappa} (i\ket{0}\bra{0}_k + \ket{1}\bra{1}_k) \otimes \mathbf{I}_{\text{others}}$ for $k = 1, \ldots, n$, and $\kappa$ represents the strength of the noise. Additionally, we consider a systematic error Hamiltonian $H^{(1)} = \kappa \beta \sum_{j=1}^n Z_j$. In this study, we do not consider measurement error and statistical error due to finite shots.

We randomly choose 100 pairs of eigen-states $(a,b)$ to demonstrate our error mitigation strategies in statistics.
For each pair of $(a,b)$, we prepare the initial state $\frac{1}{\sqrt{2}}(\ket{\phi_a}+\ket{\phi_b})$, evolve the state with noise and obtain expectation values $\langle 2 \ketbra{\phi_b}{\phi_a} \rangle (k\Delta T)$ where $k=0,1,\cdots,L-1$.
The ideal results of eigen-energy differences $E_{ba}=E_b-E_a$ are given by exact diagonalization.
To improve the estimation accuracy, we use the matrix pencil method \cite{sarkarUsingMatrixPencil1995} instead of discrete Fourier transform to retrieve the oscillation frequency (see ``Methods'') from the signal $\{\langle 2 \ketbra{\phi_b}{\phi_a} \rangle (k\Delta T)\}$ and estimate $E_{ba}$.
We note that the Hamiltonian reshaping and Hamiltonian rescaling methods also work well when the initial state contains multiple eigenstates (see Supplementary Information).

First, we demonstrate the impact of the Hamiltonian reshaping error mitigation strategy. We apply 100 random Pauli operations to generate reshaped Hamiltonians and average their eigen-energy estimates to achieve error mitigation. Considering local error in our settings, we also assess the error mitigation effect using reshaped Hamiltonians with a smaller set $\{I^{\otimes n}, X^{\otimes n}, Y^{\otimes n}, Z^{\otimes n}\}$. The mitigated relative estimation error and the distribution of relative error for each eigenvalue difference are presented in Fig.~\ref{fig::randomPauli}. Our findings indicate that this strategy efficiently reduces the relative estimation error statistically.

Secondly, we demonstrate the effectiveness of Hamiltonian rescaling in mitigating errors. The results presented in Fig.~\ref{fig::rescaling} show that our error mitigation strategy is effective under conditions of weak noise, with the second-order correction providing better mitigation than the first-order correction.
When the noise is strong, the mitigated error may exceed the unmitigated error because perturbation theory breaks down in this regime, rendering the error mitigation methods ineffective. In the strong noise regime, the first-order method can outperform the second-order estimate because experiments with a large rescaling factor $c$ may fall outside the perturbative regime, where perturbation theory is less effective, while experiments with smaller rescaling factors remain within a regime where perturbation theory still holds.

\begin{figure*}[t!]
    \centering
    \includegraphics[width=\textwidth]{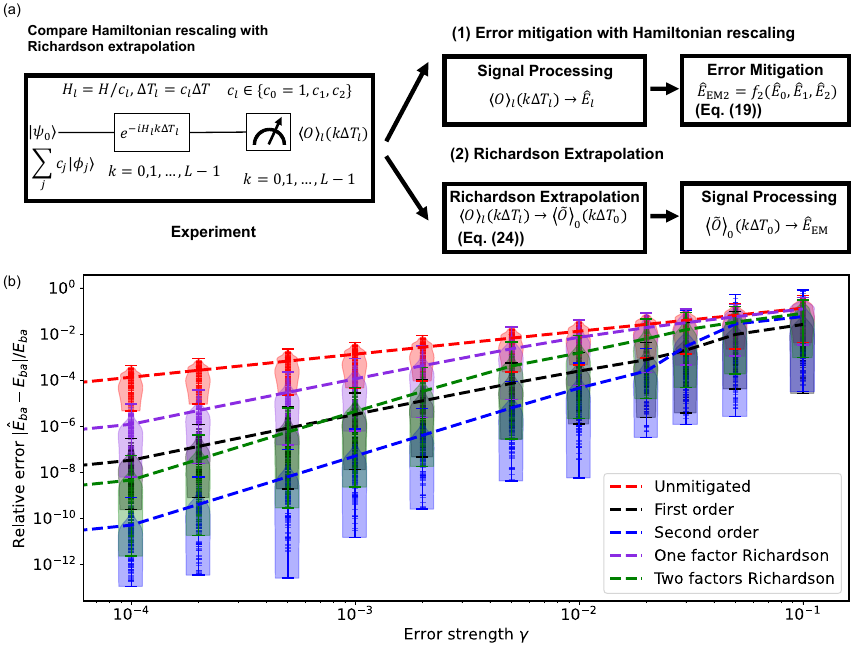}
    \caption{
    Comparison of Hamiltonian Rescaling Strategy with Richardson Extrapolation \cite{temmeErrorMitigationShortDepth2017}. Although both the Hamiltonian rescaling strategy and Richardson extrapolation utilize the same experimental data, they differ fundamentally in their approach to error mitigation. (a)(1) In our Hamiltonian rescaling strategy, the signal from the experimental data is first processed to estimate the eigenvalue differences, followed by the application of the error mitigation procedure. (a)(2) In contrast, the Richardson extrapolation method mitigates the signal corresponding to the mean value of the measurement observable before estimating the eigenvalue differences using the mitigated signal. (b) Numerical comparison of these two strategies under the conditions described in Fig. \ref{fig::rescaling}. Given that the $\langle O \rangle$ vs. $t$ signal behaves as a summation of damped oscillation modes, as discussed in ``Results'', our Hamiltonian rescaling strategy demonstrates higher efficiency compared to the standard Richardson extrapolation, as shown in panel (b).
    We note that the sample complexity of the Hamiltonian rescaling strategy is slightly higher than that of the Richardson extrapolation method (see Supplementary Information).
    }
    \label{fig::CompareWithRE}
\end{figure*}

It has been observed that more efficient methods for Hamiltonian reshaping can be developed for specific types of noise, and we numerically evaluate such a method here. When the dominant noise type in the quantum simulator is identified, it becomes feasible to design a reduced set $G$ within the Hamiltonian reshaping strategy, optimized to counteract this particular noise. For example, if the stochastic noise is amplitude damping with Lindblad operators $L_k=\sqrt{\kappa} \ket{0}\bra{1}_k \otimes \mathbf{I}_{\text{others}}$, for $k=1, \ldots, n$, and the systematic error in the Hamiltonian is represented by $H^{(1)}=\kappa \beta \sum_{j=1}^n Z_j$, then setting $G=\{\mathbf{I}^{\otimes n}, X^{\otimes n}\}$ is advantageous. This configuration is justified because $L_k^\dagger=X^{\otimes n} L_k X^{\otimes n}$, and $\{H^{(1)},X^{\otimes n}\}=0$ in this scenario. We test the tailored strategy with a transverse field $XX$ model:
\begin{equation}
    H=-g \sum_{j=1}^{n-1} X_j X_{j+1} - \sum_{j=1}^n X_j - \sum_{j=1}^n Y_j.
\end{equation}
The results in  Fig.~\ref{fig::T1noise} show that the Hamiltonian reshaping strategy tailored for the specific yet practical type of noise can effectively mitigate the error.

We also compare the mitigation effect of Hamiltonian rescaling strategy with Richardson extrapolation method \cite{temmeErrorMitigationShortDepth2017} in Fig.~\ref{fig::CompareWithRE}. The standard Richardson extrapolation method mitigate the estimation error of the expectation value of observable. First, with the noisy estimates $\langle O \rangle_l (k\Delta T_l)$ from rescaling factors $c_l$, one can get error mitigated expectation value of observable $\langle \tilde{O} \rangle_0 (k\Delta T_0)$ at each time. Then one can retrieve the energy difference with the mitigated observable by matrix pencil method. The mitigated estimate of observable expectation value is given by
\begin{equation}
    \langle \tilde{O} \rangle_0 (k\Delta T_0)=\frac{1}{c_1-1}(c_1 \langle O \rangle_0 (k\Delta T_0) - \langle O \rangle_1 (k\Delta T_1))
\end{equation}
with single rescaling factor $c_1$ and
\begin{align}
    \langle \tilde{O} \rangle_0 (k\Delta T_0)=\frac{c_1 c_2}{(c_1-1)(c_2-1)}\langle O \rangle_0 (k\Delta T_0) \nonumber\\+ \frac{c_2}{(c_1-c_2)(c_1-1)}\langle O \rangle_1 (k\Delta T_1) \nonumber\\ - \frac{c_1}{(c_2-1)(c_1-c_2)} \langle O \rangle_2 (k\Delta T_2)
\end{align}
with two rescaling factors $c_1$ and $c_2$. The numerical results indicate that our Hamiltonian rescaling strategy outperforms the standard Richardson extrapolation method, as anticipated before.

\begin{nolinenumbers}
\begin{table*}[t!]
    \centering
    \begin{tabular}{c|c|c}
        \toprule
         & Hamiltonian Reshaping & Hamiltonian Rescaling\\
        \midrule
        Total sample times $\begin{aligned} N_T \end{aligned}$ &
        $\begin{aligned}
            &N_T \sim \frac{(f(d_{ab} |\widetilde{\mathcal{D}}| T))^2 N_m^2}{T^2 ((\sigma_{E_{ba}}^{(\mathrm{EM})})^2-C |\widetilde{\mathcal{D}}|^2/N_P)}\\
            &\text{(randomly choosing reshaping operation)}\\
            &N_T \sim \frac{(f(d_{ab} |\widetilde{\mathcal{D}}| T))^2 N_m^2}{T^2 (\sigma_{E_{ba}}^{(\mathrm{EM})})^2}\\
            &\text{(average all the reshaping operation's results)}
        \end{aligned}$ &
        $\begin{aligned}
            &N_T \sim \frac{4c^2}{(c-1)^2} \frac{(f(d_{ab} |\widetilde{\mathcal{D}}| T))^2 N_m^2}{T^2 (\sigma_{E_{ba}}^{(\mathrm{EM})})^2}\\ &\text{(with single rescaling factor)}\\
            &N_T \sim 3 F(c_1,c_2)^2 \frac{(f(d_{ab} |\widetilde{\mathcal{D}}| T))^2 N_m^2}{T^2 (\sigma_{E_{ba}}^{(\mathrm{EM})})^2}\\ &\text{(with two rescaling factors)}
        \end{aligned}$
        \\
        \midrule
        $\begin{aligned} \text{When}~d_{ab} |\widetilde{\mathcal{D}}| T \ll 1 \end{aligned}$ & 
        \multicolumn{2}{c}{$\begin{aligned}
            N_T=\mathcal{O}(N_m^2 T^{-2}(1+\mathrm{poly}(|\widetilde{\mathcal{D}}|T)))
        \end{aligned}$}
        \\
        \midrule
        $\begin{aligned} \text{When}~d_{ab} |\widetilde{\mathcal{D}}| T \gg 1 \end{aligned}$ &
        \multicolumn{2}{c}{$\begin{aligned}
            N_T=\mathcal{O}(N_m^2 |\widetilde{\mathcal{D}}|^3 T)
        \end{aligned}$}
        \\
        \bottomrule
    \end{tabular}
    \caption{Complexity analysis of our error mitigation strategies. We give the the total sample times $N_T$ of Hamiltonian reshaping and Hamiltonian rescaling strategy when the error strength $|\widetilde{\mathcal{D}}|$ is small in the first row, analyze the asymptotic behavior of them when $d_{ab} |\widetilde{\mathcal{D}}| T \ll 1$ in the second row and when $d_{ab} |\widetilde{\mathcal{D}}| T \gg 1$ in the third row. Here $f(d_{ab} |\widetilde{\mathcal{D}}| T)$ and $F(c_1,c_2)$ are given in Supplementary Information. $C=N_P \text{var} (\mathrm{Im} \bra{\phi_a}U^\dagger \widetilde{\mathcal{D}} [U\ketbra{\phi_a}{\phi_b}U^\dagger] U \ket{\phi_b})/|\widetilde{\mathcal{D}}|^2$, $d_{ab}=-\frac{\mathrm{Re} \bra{\phi_a} \widetilde{\mathcal{D}} [\ketbra{\phi_a}{\phi_b}] \ket{\phi_b}}{|\widetilde{\mathcal{D}}|}>0$ are constant. $T=L\Delta T$ is the time scale of the maximum evolution time of single experiment (which is proportional to the circuit depth in the language of digital quantum simulation). $N_m$ is the number of damped oscillation modes in the signal which is constant when noise strength is small, but will grows exponentially as the number of qubits $n$ grows when noise is not small such that noisy modes should be considered. $N_P$ is the number of Pauli operations to reshape the Hamiltonian. $\sigma_{E_{ba}}^{(\mathrm{EM})}$ is the standard deviation of the mitigated eigen-energy difference and we fix it to the precision we want to obtain the sample complexity. We note that these results only works when noise strength $|\widetilde{\mathcal{D}}|$ is small and the sample complexity will grows exponentially when $|\widetilde{\mathcal{D}}|$ is large as $N_T=\mathcal{O}(N_m^2 L e^{2 d_{ab} |\widetilde{\mathcal{D}}| \Delta T})$.}
    \label{tab:ComplexityAnalysis}
\end{table*}
\end{nolinenumbers}

\textbf{{\em Complexity analysis.}}
The sample complexity of our error mitigation strategies can be obtained by analyzing the uncertainty in the mitigated eigen-energy estimation, which is intrinsically linked to the unmitigated estimates of eigen-energy difference. This relationship is fundamentally governed by the Fourier analysis underlying the many-body spectroscopy protocol.
When employing a discrete Fourier transform to analyze the Fourier spectrum, the precision of the frequency determination is inherently limited by $2\pi/L$. Consequently, the uncertainty in the energy difference is bounded by $2\pi/(L\Delta T)$.

To refine our frequency analysis, we consider the matrix pencil method and the least squares regression method to accurately retrieve Fourier frequencies. Both methods exhibit similar sample complexity, and we will employ least squares regression to analyze this complexity. Let $y_k = \langle O \rangle (k \Delta T)$ for $k = 0, 1, \ldots, L-1$, and denote $\omega_j$ as the Fourier frequencies corresponding to the list ${\langle O \rangle (k \Delta T)}$. Assuming the relationship between $y_k$ and $k$ can be modeled as a sum of damped oscillation modes, we express this as: $\hat{y}_k=\sum_{j=1}^{M} C_j r_j^k e^{i\omega_j k}$, where $C_j$, $r_j$, and $\omega_j$ are parameters to be determined by regression, and $M$ is the number of significant modes discerned, ideally from the discrete Fourier transform results. The accuracy of this model increases with $L$, recommending $L \gg N_m$ (where $N_m$ denotes the number of modes involved) and at least $L \geq N_m$.
The optimization is performed over the cost function: $\ell=\sum_{k=0}^{L-1} N_k |y_k-\hat{y}_k|^2$ where $N_k$ indicates the number of repetitions of the quantum circuit to acquire $y_k$. Optimization seeks to minimize the multi-variable function $\ell (C_j,r_j,\omega_j)$ to obtain optimal $\omega_j$ so that $\delta \phi_{ab}/\Delta T$ can be retrieved by $\omega_j/\Delta T$.
The regression process can be executed using any multi-variable optimization technique on a classical computer. While the cost function is typically non-convex, we can initialize the parameters using results obtained from discrete Fourier transform or other signal processing techniques that identify Fourier frequencies.

The uncertainty associated with this method, as well as the estimation of sample complexity for the many-body spectroscopy protocol, are derived similarly to the approach detailed in Appendix D3 of Ref.~\cite{neillAccuratelyComputingElectronic2021}. For further elaboration, readers can refer to Supplementary Information, which provides a derivation.
The relationship between the uncertainty of the phase estimation result and the noise strength $|\widetilde{\mathcal{D}}|$ is determined by
\begin{widetext}
    \begin{align}
    \frac{1}{\sigma_{\omega_\alpha}^2} \sim 2N_k C_\alpha C_\alpha^* \frac{L^2 r_\alpha^{2 L}-2 L^2 r_\alpha^{2 L+2}+L^2 r_\alpha^{2 L+4}+2 L r_\alpha^{2 L+2}+r_\alpha^{2 L+2}-2 L
   r_\alpha^{2 L+4}+r_\alpha^{2 L+4}-r_\alpha^4-r_\alpha^2}{\left(r_\alpha^2-1\right){}^3}
   \label{eqn::ComplexityOfPhaseEstimation}
\end{align}
\end{widetext}
where $\sigma_{\omega_\alpha}$ is the standard deviation of estimation result of the damped oscillation frequencies which characterize the uncertainty and $r_\alpha \sim e^{\mathrm{Re} \bra{\phi_a} \widetilde{\mathcal{D}} [\ketbra{\phi_a}{\phi_b}] \ket{\phi_b} \Delta T}=e^{-d_{ab} |\widetilde{\mathcal{D}}| \Delta T}$ from Eq.~\eqref{eqn::DampedOscillation} (where $d_{ab}=-\mathrm{Re} \bra{\phi_a} \widetilde{\mathcal{D}} [\ketbra{\phi_a}{\phi_b}] \ket{\phi_b}/|\widetilde{\mathcal{D}}|>0$).
When $|\widetilde{\mathcal{D}}|$ is large such that $r_\alpha \ll 1$, we obtain
\begin{align}
    \frac{1}{\sigma_{\omega_\alpha}^2} \sim 2N_k C_\alpha C_\alpha^* r_\alpha^2 = 2N_k C_\alpha C_\alpha^* e^{-2 d_{ab} |\widetilde{\mathcal{D}}| \Delta T}.
\end{align}
This analysis demonstrates that the variance of the phase estimation protocol will increase exponentially with the noise strength $|\widetilde{\mathcal{D}}|$, which is usually proportional to the number of qubits $n$. Consequently, we can determine the requisite sample complexity to achieve a specified precision. This is formally articulated in our results, detailed in Supplementary Information, which establish that the total number of required samples (or total sample times), $N_T$, is given by:
\begin{align}
    N_T=\mathcal{O}(N_m^2 L e^{2 d_{ab} |\widetilde{\mathcal{D}}| \Delta T}).
\end{align}
Here, we assume $C_\alpha \sim 1/N_m$  and $N_m$ represents the number of modes in the considered signal, which remains constant for low noise strengths as it is determined by the number of eigenstates included in the initial state and is independent of the number of qubits $n$. However, $N_m$ grows exponentially with increasing $n$ when the noise strength is substantial, necessitating the consideration of noisy damped oscillation modes. While perturbation theory is applicable when the noise strength is minimal, these findings highlight an inherent exponential growth in sample complexity within our error mitigation strategies, similar to all other error mitigation methods.

Our analysis utilizes perturbation theory and thus is limited to scenarios where the noise strength is sufficiently low, specifically when $d_{ab} |\widetilde{\mathcal{D}}| \Delta T \ll 1$ and $r_\alpha \rightarrow 1^-$. Under these conditions, we have established (see Supplementary Information) that the total sample times $N_T$ required by our error mitigation strategies can be accurately predicted using the formulas listed in Table \ref{tab:ComplexityAnalysis}.
Furthermore, the principles underlying our analysis are applicable to other methods for retrieving oscillation frequencies, as these techniques essentially process the same experimental data through classical means.
\\

\textbf{\large{{Discussion:}}}

In this work, we have developed two error mitigation strategies tailored to many-body spectroscopy technique for solving eigen-energy difference estimation problems using analog quantum simulators. These strategies are grounded in perturbation theory, with the Lindblad master equation providing the theoretical framework. We also analyze the computational complexity of these strategies.
Our numerical experiments demonstrate the effectiveness of the proposed methods, suggesting that these strategies could be practically implemented on current quantum hardware.

Compared with our Hamiltonian rescaling strategy, our Hamiltonian reshaping strategy has the benefits below.
First, the Hamiltonian reshaping strategy is easier to implement on real quantum devices, when dealing with simpler noise types such as local noise. This is because it does not require control over the amplitude of the Hamiltonian, and the reshaped Hamiltonians only involve changing the sign of local Pauli terms, which is significantly simpler than altering the sign of interaction terms, as discussed in ``Results''.
Second, we can prove that less total sample times are needed with Hamiltonian reshaping strategy when we mitigate the error with all the unitary operators in the reshaping operator set $G$ so that we don't consider the statistical error caused by randomly choosing reshaping operator $U \in G$ (see Supplementary Information).

Our Hamiltonian rescaling strategy also offers several advantages. First, the Hamiltonian rescaling strategy can achieve more accurate results, as it is capable of mitigating errors to the second order, whereas the Hamiltonian reshaping strategy is primarily based on first-order correction. Second, the Hamiltonian rescaling strategy is easier to implement on real quantum devices when the noise type is complex, as this would require a large set $G$ for reshaping the Hamiltonian in the Hamiltonian reshaping approach.

This study provides several important insights for future research. Firstly, the many-body spectroscopy protocol and the associated error mitigation strategies could be experimentally validated on real quantum devices.
These methods can be implemented for the problems which can be mapped to spin system, while further research is needed for the error mitigation method with simulation problems on fermionic system directly, such as the quantum chemistry problems.
Secondly, our findings indicate that the relationship between the observable expectation value $\langle O \rangle$ and the error strength is characterized by a sum of damped oscillation modes, as opposed to the linear, polynomial, or exponential dependencies commonly assumed in previous studies on zero-noise extrapolation. This observation underscores the need for developing problem-specific error mitigation techniques in the NISQ era, enabling us to leverage the specific properties of the problem to simplify experimental implementation and enhance the effectiveness of error mitigation.
\\

\textbf{\large{{Methods:}}}

\textbf{Proof of the Hamiltonian reshaping strategy.}
We prove that the error mitigation property of the Hamiltonian reshaping, i.e. Eq.~\eqref{eqn::condition}, is satisfied if the random unitary set $G$ is the set of $n$ qubits' Pauli operations.
\begin{proof}
Consider the noisy super-operator $\widetilde{\mathcal{D}}[\cdot]=-i[H^{(1)},\cdot]+\mathcal{D}[\cdot]$, which is composed of two type of noise: the unitary noise $-i[H^{(1)},\cdot]$ and stochastic noise 
$\mathcal{D}[\cdot]=\frac{1}{2} \sum_k (2 L_k \cdot L_k^\dag - L_k^\dag L_k \cdot - \cdot L_k^\dag L_k)$.

For the unitary noise we have the property $\tr{H^{(1)}}=0$. The contribution of unitary noise for the phase estimation after Hamiltonian reshaping is
    \begin{eqnarray}
        &&\quad \overline{\bra{\phi_a}U^\dagger(-i[H^{(1)},U\ketbra{\phi_a}{\phi_b}U^\dagger])U\ket{\phi_b}} \nonumber \\
        &&=-i(\overline{\bra{\phi_a}U^\dagger H^{(1)} U \ket{\phi_a}}-\overline{\bra{\phi_b} U^\dagger H^{(1)} U\ket{\phi_b}})\nonumber \\
        &&=0 
    \end{eqnarray}
    where $U$s are random Pauli operations. Here we use the property that Pauli group forms unitary 1-design
    \begin{eqnarray}
        \overline{U^\dagger O U}=\tr{O}\,\frac{I}{d}
    \end{eqnarray}
    where $O$ is an arbitrary operator, $I$ is the identity operator and $d$ is the system dimension.
    
    Then we need to prove the contribution from stochastic noise $\mathcal{D}[\cdot]$ for the phase estimation is also zero. 
    Since
    \begin{equation}
        \overline{\mathrm{Im} \bra{\phi_a}U^\dagger L_k^\dagger L_k U \ket{\phi_a}}=\overline{\mathrm{Im} \bra{\phi_b} U^\dagger L_k^\dagger L_k U \ket{\phi_b}}=0
    \end{equation}
    we just need to prove
    \begin{equation}
        \overline{\mathrm{Im} \sum_k \bra{\phi_a}U^\dagger L_k U\ketbra{\phi_a}{\phi_b}U^\dagger L_k^\dagger U \ket{\phi_b}}=0\,.
    \end{equation}
Lindblad operators $L_k$s can be expanded into Pauli operators 
    \begin{equation}
        L_k=\frac{1}{2^n} \sum_{j} \tr (L_k P_j) P_j
    \end{equation}
    where $P_j$s are Pauli operators, and by which we have
    \begin{eqnarray}
        \mathrm{Im} \sum_k \bra{\phi_a}U^\dagger L_k U\ketbra{\phi_a}{\phi_b}U^\dagger L_k^\dagger U \ket{\phi_b} \nonumber  \\
        = \frac{1}{4^n} \mathrm{Im} \sum_k \sum_{j_1} \sum_{j_2} \tr (L_k P_{j_1}) \tr (L_k^\dagger P_{j_2})  \nonumber \\ \cdot \bra{\phi_a}U^\dagger P_{j_1} U\ketbra{\phi_a}{\phi_b}U^\dagger P_{j_2} U \ket{\phi_b}\,.
    \end{eqnarray}
    As $U$ are chosen uniformly randomly from $n$ qubits' Pauli operations, the averaged contribution of stochastic noise is
    \begin{eqnarray} \label{eq:ave sto contribution}
       && \overline{\mathrm{Im} \sum_k \bra{\phi_a}U^\dagger L_k U\ketbra{\phi_a}{\phi_b}U^\dagger L_k^\dagger U \ket{\phi_b}} \nonumber \\
        &=& \frac{1}{4^n} \mathrm{Im} \sum_k \sum_{j_1} \sum_{j_2} \tr (L_k P_{j_1}) \tr (L_k^\dagger P_{j_2}) \nonumber \\
        && \cdot \, \overline{\bra{\phi_a}U^\dagger P_{j_1} U\ketbra{\phi_a}{\phi_b}U^\dagger P_{j_2} U \ket{\phi_b}} \nonumber\\
        &=& \frac{1}{4^n} \mathrm{Im} \sum_k \sum_{j_1} \sum_{j_2} \tr (L_k P_{j_1}) \tr (L_k^\dagger P_{j_2}) \nonumber \\ 
        && \cdot \, \bra{\phi_a} P_{j_1} \ketbra{\phi_a}{\phi_b} P_{j_2} \ket{\phi_b} \overline{\zeta(P_{j_1},U) \zeta(P_{j_2},U)}
    \end{eqnarray}
    where
    \begin{equation}
        \zeta(P,U)=\left\{ \begin{array}{l}
            1, \quad \text{if} \quad [P,U]=0, \\ -1, \quad \text{if} \quad \{P,U\}=0.
        \end{array} \right.
    \end{equation}
    As
    \begin{equation}
        \overline{\zeta(P_{j_1},U) \zeta(P_{j_2},U)} = \overline{\zeta(P_{j_1}P_{j_2},U)}
    \end{equation}
    since $\zeta(P_{j_1},U) \zeta(P_{j_2},U) = \zeta(P_{j_1}P_{j_2},U)$ when $U$ is Pauli operator which has been proved in the supplementary material in \cite{caiConstructingSmallerPauli2019},
    and $P_{j_1}P_{j_2} \neq \mathbf{I}^{\otimes n}$ when $j_1 \neq j_2$, we get $\overline{\zeta(P_{j_1}P_{j_2},U)}=0$ for $j_1 \neq j_2$.  The summation in Eq.~(\ref{eq:ave sto contribution}) only includes the terms which satisfy $j_1=j_2$ so that
    \begin{eqnarray}
       && \overline{\mathrm{Im} \sum_k \bra{\phi_a}U^\dagger L_k U\ketbra{\phi_a}{\phi_b}U^\dagger L_k^\dagger U \ket{\phi_b}} \nonumber\\
        &=& \frac{1}{4^n} \mathrm{Im} \sum_k \sum_{j_1} |\tr (L_k P_{j_1})|^2  \nonumber \\ 
        &&\quad \cdot \bra{\phi_a} P_{j_1} \ketbra{\phi_a}{\phi_b} P_{j_1} \ket{\phi_b} \overline{\zeta(\mathbf{I}^{\otimes n},U)}.
    \end{eqnarray}
    Because the Pauli operations are Hermitian, we get
    \begin{equation}
        |\tr (L_k P_{j_1})|^2 \bra{\phi_a} P_{j_1} \ketbra{\phi_a}{\phi_b} P_{j_1} \ket{\phi_b} \overline{\zeta(\mathbf{I}^{\otimes n},U)} \in \mathbb{R}\,.
    \end{equation}
    Finally, we prove 
    \begin{equation}
        \overline{\mathrm{Im} \sum_k \bra{\phi_a}U^\dagger L_k U\ketbra{\phi_a}{\phi_b}U^\dagger L_k^\dagger U \ket{\phi_b}}=0.
    \end{equation}
\end{proof}

We also prove that Eq.~\eqref{eqn::condition} is satisfied when local noise and smaller set of Pauli operators are considered.

\begin{proof}
    When the unitary noise and stochastic noise are local, we can replace the random Pauli operations $U$ with random local Pauli operations' tensor product in the proof above, expand local Lindblad operators $L_k$s into local Pauli operators instead, then the proof above still works when $G$ is the set of local Pauli operations' tensor product $\{I^{\otimes n}, X^{\otimes n}, Y^{\otimes n}, Z^{\otimes n}\}$.

    When the unitary noise and stochastic noise have specific properties, it's possible to find smaller $G$. For example, when $L_k=\sqrt{\kappa} \left( (0,1),(0,0)\right)_k \otimes \mathbf{I}_{\text{others}}$ (for $k=1, \ldots, n$), and the systematic error in the Hamiltonian is represented by $H^{(1)}=\kappa \beta \sum_{j=1}^n Z_j$, then setting $G=\{\mathbf{I}^{\otimes n}, X^{\otimes n}\}$ also works and here is the proof.
    
    As $I^{\otimes n} H^{(1)} I^{\otimes n} + X^{\otimes n} H^{(1)} X^{\otimes n}=0$, the contribution of unitary noise for the phase estimation after Hamiltonian reshaping is $0$, although here $G=\{\mathbf{I}^{\otimes n}, X^{\otimes n}\}$ doesn't form unitary 1-design.
    And as $L_k^\dagger=X^{\otimes n} L_k X^{\otimes n}$,
    \begin{eqnarray}
        &&\overline{\mathrm{Im} \sum_k \bra{\phi_a}U^\dagger L_k U\ketbra{\phi_a}{\phi_b}U^\dagger L_k^\dagger U \ket{\phi_b}}\nonumber\\
        &=&\mathrm{Im} \sum_k ( \bra{\phi_a} L_k \ketbra{\phi_a}{\phi_b} L_k^\dagger  \ket{\phi_b} \nonumber\\ 
        &&+ \bra{\phi_a} L_k^\dagger \ketbra{\phi_a}{\phi_b} L_k \ket{\phi_b})=0
    \end{eqnarray}
    so that the stochastic noise's contribution is also $0$.
\end{proof}

\textbf{First-order degenerate perturbation.}
We show the result of first-order degenerate perturbation and discuss the influence on our work.

Consider the Liouvillian operator $\mathcal{L}$ which has $p$ degenerate eigenvectors $\ketbra{\phi_{a_1}}{\phi_{b_1}}$, $\cdots$, $\ketbra{\phi_{a_p}}{\phi_{b_p}}$ when there is no noise and the corresponding eigenvalue is $\Lambda_{ab}$.
When weak noise $\widetilde{\mathcal{D}}[\cdot]$ is applied, the left eigenvectors can be considered the same as the right eigenvectors of $\mathcal{L}$, and the eigenvalues will be split because of noise.
The eigenvalues after first-order degenerate perturbation are $\Lambda_{ab} + \Delta \Lambda_{abj}$ where $j=1,2,\cdots,p$.
$\Delta \Lambda_{abj}$ is equal to the eigenvalues of the matrix $\widetilde{D}_{mn}=\bra{\phi_{a_m}}\widetilde{\mathcal{D}}[\ketbra{\phi_{a_n}}{\phi_{b_n}}]\ket{\phi_{b_m}}$ where $m,n=1,2,\cdots,p$.

Next, we discuss the result in detail.
Consider the noise effect on a specific $\ketbra{\phi_{a_l}}{\phi_{b_l}}$.
Noise will affect the estimation result by the protocol unless the eigenvalues of $\bra{\phi_{a_m}}\widetilde{\mathcal{D}}[\ketbra{\phi_{a_n}}{\phi_{b_n}}]\ket{\phi_{b_m}}$ which correspond to the degenerated subspace that $\ketbra{\phi_{a_l}}{\phi_{b_l}}$ belongs to are real (but in most cases this requirement cannot be satisfied).
Small energy differences of the Hamiltonian are hard to discriminate as all the $\ketbra{\phi_j}{\phi_j}$ are degenerate which corresponds to the zero energy gap and will be split into eigenvalues near zero caused by the first-order degenerate perturbation theory.
And it's possible to affect the protocol to retrieve eigenvalues that correspond to degenerate eigenstates.
However, this energy splitting effect is independent of the strength of the Hamiltonian so that it can be canceled by our Hamiltonian rescaling strategy since our strategy only uses the difference of $\delta \phi_{ab}/\Delta T$ in Eq.~\eqref{eqn::First-Order} and Eq.~\eqref{eqn::secondOrderResult}.
We note that our error mitigation strategies will fail if the noise is strong so that the split peaks overlap with peaks that correspond to other eigenvalues.

\textbf{Methods to retrieve frequencies.}
We introduce various methods to retrieve the frequencies $\omega_j$s of the summation of damped oscillation modes $y_k = \sum_{j=1}^{N_m} C_j r_j^k e^{i\omega_j k}$ where $k=0,1,2,...,L-1$.

First, we introduce the matrix pencil method which is developed and discussed in detail in \cite{sarkarUsingMatrixPencil1995}.
This method can handle noise in the signal based on singular value decomposition.
Denote
\begin{equation}
    Y=\left[\begin{array}{cccc}
        y_0 & y_1 & \cdots & y_{L_P} \\
        y_1 & y_2 & \cdots & y_{L_P+1} \\
        \vdots & \vdots & & \vdots \\
        y_{L-L_P-1} & y_{L-L_P} & \cdots & y_{L-1}
        \end{array}\right]_{(L-L_P) \times(L_P+1)}.
\end{equation}
Here $L_P$ is the pencil parameter which is better to be set between 1/2 and 2/3 (or 1/3 to 1/2) of $L$.
We apply singular value decomposition to matrix $Y$ so that
\begin{equation}
    Y=U\Sigma V^\dagger.
\end{equation}
Next we pick $M$ dominant modes with a cutoff ratio.
$M$ is the number of singular values which are larger than $\sigma_{\max}\cdot \text{cutoff}$ since $y_k$ is not accurate and noisy singular values can be dropped out in this way.
Then we consider the filtered matrix $V^\prime$ which only contains $M$ dominant right-singular vectors of $V$:
\begin{equation}
    V^\prime=[v_1,v_2,\cdots,v_M].
\end{equation}
Denote matrix
\begin{equation}
    V_1^\prime = [v_1,v_2,\cdots,v_{M-1}]
\end{equation}
and matrix
\begin{equation}
    V_2^\prime = [v_2,\cdots,v_{M}],
\end{equation}
$z_j=r_j e^{i \omega_j}$ can be retrieved by calculating the complex conjugate of the eigenvalues of $(V_1^\dagger)^+ V_2^\dagger$ where $(V_1^\dagger)^+$ is the pseudo inverse of $V_1^\dagger$.

We note that the cutoff ratio may affect the frequencies retrieving result since more noisy modes are considered as cutoff ratio decreases.
In our numerical experiments, we set the cutoff ratio as $10^{-10}$ for Hamitonian reshaping strategy in Fig.~\ref{fig::randomPauli} and Fig.~\ref{fig::T1noise}.
For Hamiltonian rescaling strategy, we set the cutoff ratio as $10^{-2}$.

The detailed parameter settings are included in our code \cite{github}.
The computational complexity of this method is limited by the size $\sim(O(L)\times O(L))$ matrix's operations.
And as it's better to set $L\gg N_m$ and at least $L \geq N_m$, the complexity is larger than the size $\sim(O(N_m)\times O(N_m))$ matrix's operations.

Next, we derive the result given by discrete Fourier transform.
We prove that the module of the spectrum after discrete Fourier transform includes the information of $\omega_j$.
The peaks of the spectrum's horizontal coordinates are $\omega_j$s.
\begin{proof}
    The Fourier spectrum is proportional to
    \begin{align}
        \mathcal{F}(\omega)=\sum_j C_j (\frac{1}{i(\omega-\omega_j+i\ln r_j)} \nonumber \\ +2\pi \delta(\omega-\omega_j+i\ln r_j)).
    \end{align}
    As $0 < r_j < 1$, the delta function is always equal to zero so that
    \begin{align}
        \mathcal{F}(\omega)=\sum_j C_j \frac{1}{i(\omega-\omega_j+i\ln r_j)}.
    \end{align}
    When $\omega=\omega_j$, both $|\mathcal{F}(\omega)|$ and $\mathrm{Re} \mathcal{F}(\omega)$ approximately reach the maximum since $r_j\rightarrow 1$ as noise is weak, and the results are approximately proportional to $-1/\ln r_j$.
\end{proof}
The computational complexity of the discrete Fourier transform is determined by a single matrix multiplication operation and it's easier than the matrix pencil method.
However, the precision of the peaks' horizontal coordinates is limited by $\frac{2\pi}{L}$ so that it can only give a rough result.

The method based on regression which is introduced in ``Results'' can improve the precision.
However, the optimization process on a $3N_m$ parameters' non-convex cost function is hard.
A possible solution is to use the result given by other methods to set appropriate initial values of the parameters, but this method is also unstable since it's possible to be optimized into local minimums.

Comparing the various methods mentioned above, we choose the matrix pencil method to retrieve the frequencies in this manuscript since the result given by the method is stable and accurate and the computational complexity is acceptable.

\begin{acknowledgments} 
\textbf{{\em Acknowledgments.}} Authors thank Shuo Yang, Yunzhe Zheng, Qinghong Yang, and Qiujiang Guo for useful discussions.
This work is supported by National Natural Science Foundation of China (Grants No. 92365111), Beijing Natural Science Foundation (Grants No. Z220002) and the Innovation Program for Quantum Science and Technology (Grant No. 2021ZD0302400).

\textbf{{\em Author contribution.}} R.-C. G. and Y. G. developed the error mitigation strategies and did the analytical calculations, R.-C. G. did the numerical simulation, D.E.L. initialized and supervised the project. All the authors contribute to the result analysis and manuscript preparation.
\end{acknowledgments}

\textbf{\em{{Competing interests.}}}
The authors declare no competing interests.

\textbf{{\em Data availability.}}
The simulated data is available upon request.

\textbf{{\em Code availability.}}
The source code for the numerical simulations is available at GitHub repository \cite{github}.

\bibliographystyle{apsrev4-1-titles} 

\end{document}